\documentclass[11pt]{article}

\pdfoutput=1

\usepackage[bottom]{footmisc}

\usepackage{jheppub}
\usepackage{amsfonts}
\usepackage{graphicx}
\usepackage{amsmath}
\usepackage{amssymb}

\usepackage[normalem]{ulem}
\usepackage[utf8]{inputenc}

\usepackage[english]{babel}
\usepackage[autostyle]{csquotes}
\MakeOuterQuote{"}
\usepackage{indentfirst}
\usepackage[dvipsnames]{xcolor}

\title{The trans-Planckian problem\\
and gravitational interactions} 
\author{Ted Jacobson}
\affiliation{Maryland Center for Fundamental Physics,\\University of Maryland, College Park, MD 20742, USA}
\emailAdd{jacobson@umd.edu}
\date{}

\def\m{\mu}

\def\l{\lambda}

\def\beq{\begin{equation}}
\def\eeq{\end{equation}}
\def\bea{\begin{eqnarray}}
\def\eea{\end{eqnarray}}
\def\bal{\begin{align}}
\def\eal{\end{align}}
\def\nn{\nonumber}
\def\ben{\begin{enumerate}}
\def\een{\end{enumerate}}
\def\bit{\begin{itemize}}
\def\eit{\end{itemize}}
\def\edoc{\end{document}}

\def\la{\langle}
\def\ra{\rangle}
\def\e{\epsilon}
\def\d{\delta}
\def\D{\Delta}
\def\k{\kappa}
\def\l{\lambda}
\def\m{\mu}

\def\s{\sigma}
\def\o{\omega}

\def\g{\gamma}

\def\O{\Omega}

\def\half{\textstyle{\frac{1}{2}}}

\def\3halfs{\textstyle{\frac{3}{2}}}

\def\tu{\widetilde u}

\def\tP{trans-Planckian }

\abstract{The strong gravitational interaction of counter-propagating 
quantum field modes is of foundational importance to the trans-Planckian
problem of black hole horizons. This article, in
memory of Renaud Parentani, is primarily an exposition of Renaud's
attempt, using perturbation theory, a large $N$ approximation, 
and spherical reduction, to capture a mechanism by which those 
interactions might quench
the trans-Planckian near-horizon quantum field correlations. Before 
giving a detailed account of Renaud's calculation, the 
trans-Planckian problem is briefly introduced, 
and the paper concludes with a discussion of related issues and questions.}

\begin{document}

\maketitle
\flushbottom

\section{Introduction}
Renaud Parentani was one of the most insightful and creative
physicists I have known.  
His methodology was to take a question --- whether it be 
foundational or phenomenological --- and, with the help of 
simplifications, approximations, and astute technical
devices, craft it into a formulation
that he then studied with precise  analysis 
from which he drew sharp results and lessons. 
I treasured 
his instruction, collaboration and friendship, both as a physics colleague and on 
a personal level. Time spent with 
him was always illuminating, whether on the subject of physics or any of 
his many other passions.
 His passing was a terrible loss for me personally 
and for science. 

Renaud and I first met in 1995, thanks to a shared interest in 
the trans-Planckian problem of black hole horizons. Although we wrote only
one paper together on that topic \cite{Jacobson:2007jx}, it was a major theme
underlying both of our lines of research, and the subject of countless discussions.
For this contribution in his memory, I focus on his efforts
to go beyond the semiclassical approximation in order to understand the role
of quantum gravitational interactions in ``taming the trans-Planckian monster''
\cite{Brout:1995wp}.
This involves an area of his research that remained quite tentative and
incomplete, so is rather unlike most of his work. It is also, I suspect, 
relatively unknown, perhaps in part because his papers on the 
subject are not so easy to follow. 
I wanted to understand it better myself, 
and to spread the word about it, with the hope that it will bear fruit.

\section{The trans-Planckian problem}

Traced backward in time toward the horizon, 
an outgoing field mode
blueshifts exponentially, as measured
by freely falling observers.
Near the horizon 
this mode frequency increases as
$e^{\k t}$, where $\k$ is 
the surface gravity 
and $t$ is the backwards-elapsed Killing time. 
For a solar mass black hole
$\kappa\sim 10^{5}$ inverse seconds, so 
over a Killing time of one second 
the frequency increases by $\sim e^{10^5}$,
quickly surpassing exponentially far beyond the Planck frequency.
These modes are populated at 
the Hawking temperature $\hbar\k/2\pi$ in the
free-fall vacuum state, but they also play a classical role;
for example an outgoing mode
of the electromagnetic field 
could be excited by an oscillating electric dipole
outside the black hole.

To account for the physics of outgoing modes,
relativistic field theory on 
a fixed black hole spacetime background thus
relies on an exponentially \tP 
(as measured in any given free fall frame)
reservoir of modes at the horizon.\footnote{The earliest
use of the term ``trans-Planckian'' in this context 
that I am aware of was by
't Hooft~\cite{tHooft:1994tah}.}
Since the theory
possesses local Lorentz invariance
there is prima facie nothing wrong 
with this. However, it has the dubious
effect of forcing us
to assume local 
Lorentz invariance of the theory 
to unbounded boosts, far beyond any that could
ever be measured experimentally.
This doesn't pass the smell test. 
Moreover, that \tP reservoir would carry an 
infinite entanglement entropy, contradicting the
finiteness of the Bekenstein-Hawking black hole
entropy $A/4G\hbar$. One expects that 
quantum gravity somehow eliminates the 
apparent role of \tP degrees of freedom,
and the question is how does that work?

Another facet of the question is presented by 
exponential (or even sub-exponential) 
redshifts in cosmology, which require either
a \tP reservoir of modes in the early universe, 
or some sort of mode creation, to account for the
presence of the highest observed frequency modes 
today, even if inflation never took place.
Mode creation, that is, creation of new degrees
of freedom as the universe expands,
does not seem to me implausible. It might
resolve the cosmological \tP problem, but 
it is presumably not relevant for the black hole
one, since a (stationary) black hole exterior spacetime has 
time translation symmetry. For a review of both 
of these puzzles and related work, as of the
turn of the century, see \cite{Jacobson:1999zk}.

Renaud gave much thought to 
the \tP problem.
Most of his work in this domain involved
exploration of Hawking radiation and the related 
inflationary particle creation, using 
field theory with modified dispersion
and/or dissipation. This
original motivation also led to his 
extensive investigation of the potential impact of UV physics
on cosmological observables, as well as of classical and 
quantum properties of physical realizations
of analog black holes. 
A brief account of how he viewed these issues 
and results, as of 2002, can be found in \cite{Parentani:2002bd}.
He spent the next eighteen years further developing
these ideas. 

On the fundamental side, Renaud pursued the idea that 
quantum gravitational interactions and metric fluctuations must 
``tame'' the theory, i.e., eliminate
the apparent role of \tP field modes in Hawking radiation and 
render finite the entanglement entropy.
Although this is the least complete aspect of his work, 
I believe it is the most significant
with respect to addressing the fundamental puzzles
of horizon physics.

\section{Beyond the semiclassical description}

The conclusion section of \cite{Brout:1995wp} 
offers a colorful description of why and how quantum gravitational 
interactions might be expected to intervene and resolve
the \tP problem. Together with Barrabes and Frolov,
Renaud explored  
in \cite{Barrabes:2000fr}
the impact of a prescribed 
ensemble of black hole metric fluctuations on the 
\tP problem, 
and in several papers \cite{Parentani:2000ts,Parentani:2001tg,parentani2002towards,Parentani:2002fvb,
parentani2007beyond} 
he attempted to implement this idea 
in quantum gravity. These papers are all closely related.
They form a sequence of refinements of the initial analysis in \cite{Parentani:2000ts}.
I will base my review on the final one \cite{parentani2007beyond}, which 
exists only as an arXiv submission. The title is the same as that of 
\cite{Parentani:2002fvb}, but carries an asterisk that reads: 
``An earlier version of this work was published in Proceedings [21] in 2002. 
We postponed submitting it to the arXiv in the hope of improving the evaluation of radiative corrections. We have recently modified the text, corrected some mistakes, 
and added remarks on this difficult point which still needs work --- that hopefully someone
will take over.'' 

\subsection{The role of gravitational interactions}
\label{role}
A near-horizon 
outgoing mode, which is the exponentially blue-shifted 
ancestor of a Hawking quantum, has strong
gravitational interactions with ingoing low frequency modes once the 
invariant mass of the outgoing-ingoing pair reaches the Planck mass.
The idea Renaud pursues is that 
this interaction --- with both modes in their ground states --- obliterates 
the free-field description of the ancestor.\footnote{The earliest discussion of the essential importance of gravitational
interactions between ingoing and outgoing black hole 
modes that I am aware of was by 't Hooft~\cite{tHooft:1984kcu, tHooft:1986vqu}, 
who considered the effects 
of an ingoing ``particle'', i.e. excitation, on the 
outgoing emitted radiation.} 
That is, traced backwards in time, an outgoing mode dissipates into the 
strongly interacting 
quantum gravitational vacuum, eliminating the role of \tP frequencies in the
origin of the outgoing modes and the Hawking quanta they carry. 
To exhibit this phenomenon Renaud aims to compute 
the two-point correlation function, with one point far from the horizon and 
the other on the past light cone of that point, close to the horizon.
Within an approximation and truncation
scheme to be described below, he finds that 
the free-field light cone singularity is replaced by a finite correlation
that decays away exponentially as the point near the horizon recedes to the past.\footnote{In de Sitter spacetime an eternal inertial observer is
surrounded by a spherical horizon which behaves in many respects
like a black hole horizon~\cite{Gibbons:1977mu}, the key differences being that 
there is no asymptotically flat region and that future null infinity is spacelike
rather than null. But a de Sitter horizon presents a \tP problem as well,
and it would be interesting to adapt Renaud's model to that case.}

The reason for restricting attention to {\it low frequency} ingoing modes,
i.e., modes with wavelengths that are 
not much shorter than the Schwarzschild radius, is 
that only for them could the presence of the black hole plausibly 
make a difference. That is, the effect Renaud is trying to capture 
must arise from the {\it difference} between the flat space interactions
and those that occur in the neighborhood of the black hole horizon. 
In flat space, Planckian collisions with vacuum modes
must be occurring everywhere and all the time.
Evidently, that does not prevent, for example, an electromagnetic wave
from propagating freely through flat spacetime, since 
such a wave corresponds to 
an excitation {\it above} the interacting quantum
gravitational vacuum.

The physics is somewhat analogous to that of
the Lamb shift: as the electric charge coupling is turned on,
a free electron in the Minkowski vacuum develops an infinite self-energy
due to ``radiative corrections''.
But a physical one-electron state is an excitation above the interacting
QED vacuum, and its mass is a parameter we determine by observation,
so we do not observe the divergent effect of the radiative corrections. 
In a hydrogen atom, however, the electron is subject to slightly modified 
radiative corrections, which depend on the electron orbital.
It is the {\it difference} between these and those
for a free electron that brings about the Lamb shift, raising the
energy of the 2s${}_{1/2}$ state above that of the 2p${}_{1/2}$ state
by about one part in $10^{11}$ of the electron mass.\footnote{The difference is 
small since it depends on a high power of the small
electromagnetic coupling constant. 
In the gravitational case, the coupling grows with energy,
hence the effect of a background field might be large, 
indeed large enough to quench otherwise infinite correlations
between fields near and far from a horizon.}
Hans Bethe famously estimated the Lamb shift using a nonrelativistic
calculation, with an ad hoc  high energy cutoff at the electron mass.
Since the result depends only logarithmically on the cutoff, he was able to 
obtain a result that was quite close to the observed value.
Shortly after, fully consistent renormalization of relativistic QED 
was understood, and with that
no ad hoc cutoff is necessary. 

Similarly, Renaud finds it necessary to introduce by hand a high energy cutoff; 
and, like the Lamb shift, the result depends only logarithmically on the cutoff,
so he feels justified in drawing some tentative conclusions. But how might
one do better? Again, the analogy with the Lamb shift may be instructive.
For one thing, some kind of subtraction of the flat space gravitational 
vacuum interactions should be done, because one is perturbing around 
that vacuum. For another thing,
to master renormalization of QED it was necessary to develop a formalism 
in which Lorentz and gauge symmetry are manifest. It is plausible that,
similarly, to master the physics of the outgoing black hole modes, 
it will be essential to preserve diffeomorphism symmetry. 
Not only is that a far more complicated symmetry to handle, but 
one must also allow for different background metrics, unlike for the 
Lamb shift in which all physics takes place in Minkowski spacetime.
The challenge to improve on Renaud's calculation is daunting, 
but the potential reward is great: an account of the
quenching of near-horizon correlations
could serve as the ``Lamb shift'' of quantum 
gravity,\footnote{Albeit unfortunately lacking a precise measurement to compare to!} and bring us closer to understanding that theory.

It should be emphasized at the outset that simple
perturbation theory in the gravitational constant
$G$ will never erase a light cone singularity in a correlation 
function, so clearly something more  is needed to make headway.
Gravitational interaction becomes arbitrarily strong in the UV, 
so it is plausible that nonperturbative effects do produce erasure.
However,  to detect that, 
one would presumably need to calculate using a UV completion 
of  general relativity. It would be wonderful if AdS/CFT duality
could be deployed to this end, but that is challenging because the near horizon 
physics is difficult to capture using available holographic technology,
but perhaps future developments will make this possible. 
(See the discussion section for further comments on this.)

In any case, Renaud took a different route.
He  introduced a large number $N$ of identical scalar fields, which enables 
effects that are nonperturbative in $G$ to be calculated in a controlled way, at next to
leading order in $1/N$. In effect, the double series in $G$ and $N$ is organized into
a sum of series with the product
$GN$ held fixed, 
the $m$th series being one in powers of $G^m N$. 
The $m=1$ series yields semiclassical gravity, in which
the classical metric satisfies the semiclassical Einstein equation, 
while the $m=2$ series includes, at the next order in $1/N$, effects of fluctuations
of the energy-momentum tensor of the matter, characterized by its two-point 
correlation function. A toy model explaining how this works is discussed in 
section 9.4 of \cite{Hu:2020luk}.

I next
review some of the content of Renaud's paper  \cite{parentani2007beyond}.
My presentation will sometimes
follow a different path than his, and offer 
what may be different justifications for the approximations involved.
(I also correct a few typos and numerical factors.) 
Because his reasoning is not everywhere completely clear to me,
I may have missed or misunderstood important points, and
I may have introduced errors not present in \cite{parentani2007beyond}.
For these reasons, and since I cover only a portion, I
encourage the interested reader to consult Renaud's paper as well.

\subsection{Spherical model} 

The model consists of a massless scalar field coupled to gravity in four 
spacetime dimensions, restricted to
spherical symmetry. (Later, $N$ identical such scalar fields
will be introduced in order to exploit a systematic large $N$ approximation
when computing a correlation function.)
Restricted to spherical symmetry there are no independent
gravitational degrees of freedom, and the system is described as effectively 
two-dimensional. The background metric $g_0$ is taken to be a black hole formed
by collapse of a thin null shell with a macroscopic mass $M$ (although the details 
of how the black hole forms  play no essential role).
With spherical fluctuations included, the metric $g_0 + h$ after the black hole 
has formed
is given in one gauge by the line element\footnote{We adopt units with
$c=\hbar=1$.}
\beq\label{ds2}
ds^2 = e^\psi\left[-f\,dv^2 + 2\, dv\,dr\right] + r^2 d\Omega^2
\eeq
where 
\beq\label{f}
f= f_0 -\frac{2G\m}{r}:= 1 -\frac{2G(M +\m)}{r}\,,
\eeq
with $\psi$ and $\m$  functions of $v$ and $r$ that determine the 
deviation $h$ from the background $g_0$. 
These spherical metric fluctuations
are induced by the scalar field stress tensor.
The advanced time coordinate $v=0$ is chosen to coincide with the collapse of 
the null shell, so the line element \eqref{ds2} with $M\ne0$ is applicable 
for $v>0$ only.

The spherical mode of the four-dimensional 
scalar field  $\chi$ is described by an effective two-dimensional field,
\beq\label{phi}
\phi := \chi\sqrt{4\pi r^2}\,,
\eeq
in terms of which the action (up to total
 derivatives with respect to $v$ or $r$) is
 \beq\label{Sphi}
S^{\phi,h} = -\int dv\, dr\, \Bigl[\phi_v\phi_r  + \tfrac12 f \phi_r^2 + \tfrac12 (f_r/r)\phi^2\Bigr]\,,
\eeq
where subscripts denote partial derivative with respect to the subscript variable.\footnote{A comma will be included in the following 
when necessary to separate a derivative subscript from a previous subscript. Also, 
here and below, an expression like $\phi_r^2$ denotes the square of the derivative (not the derivative of the square).}
Note that this action depends on the metric perturbation $h$ only 
via $\mu$, i.e., $\psi$ does not enter.
The last term in the integrand of \eqref{Sphi} takes the form 
\beq\label{potential}
G\left(\frac{M+\m}{r^3} -\frac{\m_r}{r^2}\right)\phi^2\,.
\eeq
Before adding the fluctuation $\m$ in his presentation, 
Renaud drops this term on the grounds that it
produces a potential barrier that, while 
it partially scatters ingoing into outgoing modes and vice versa, 
does not affect the near horizon propagation and
is therefore not of interest for the consequences of the 
strong, in-out gravitational 
interaction.  He does not mention
the $\m$ terms that occur in \eqref{potential}.

On the static black hole background $g_0$, 
and neglecting the potential barrier term
in \eqref{Sphi}, 
the scalar field splits into ingoing and outgoing components
that are dynamically independent. 
At the classical level,
the action \eqref{Sphi} (with potential barrier term neglected)  
implies the equations of motion 
\beq\label{phi-}
(2\phi_v + f_0\phi_r)_r = 0\,.
\eeq
The solutions that satisfy $\phi_r = 0$, i.e., $\phi = \phi(v)$,
are ingoing modes, which are denoted by $\phi_+(v)$. They are constant 
on the ingoing radial null curves.
The other way to solve \eqref{phi-} is to 
satisfy 
\begin{equation}\label{nonlin}
    2\phi_v + f_0\phi_r = 0\,,
\end{equation} 
which implies that $\phi$ is constant
on the outgoing radial null curves
(cf. Appendix \ref{A}). This outgoing mode is denoted $\phi_-$, 
and is a function only of the retarded time $u$, $\phi_- = \phi_-(u)$.

The gravitational interactions are in the terms of \eqref{Sphi} involving 
$f$, namely, $\mu \phi_r^2$, $\m_r\phi^2$, and $\mu \phi^2$. 
Note that since $\phi_+$ depends only on $v$ 
in the $(v,r)$ coordinate system, $\phi_r$ involves only the outgoing
mode $\phi_-$. Moreover, whatever frequency $\phi_-(u)$ has 
with respect to $u$, its derivative with respect to $r$
diverges at the horizon (cf.~\eqref{r|v}): 
\beq\label{phirv}
\phi_r|_v = -2f_0^{-1}\phi_u|_{v}\,,
\eeq%
so the outgoing modes near the horizon have ``high local frequency''
with respect to $r$. Once $\mu$ is solved for in terms of $\phi$,
its $\phi_+$ dependence in the $\m \phi_r^2$ term 
thus entails strong gravitational interactions
between $\phi_-$ and $\phi_+$.  
Renaud's analysis tracks only this
interaction. He does not explicitly discuss why.
Perhaps he did not notice the other contributions, 
or perhaps he just chose to simplify the model by 
retaining only the dominant interaction term.
Notice that (after integration by parts on $\m_r$), 
the other two terms have the forms $\mu \phi \phi_r$
and $\mu \phi^2$, which are less singular at the horizon
since they contain only one or zero factors of $\phi_r$.
Going forward, I follow Renaud and 
drop the other terms.\footnote{Renaud explains below equation (17) 
that, since the interaction is to be computed only to lowest order in $G$, one may
use the on-shell condition $\phi_{+,r}=0$ in \eqref{Sint}, and therefore that interaction 
is governed by $\mu \phi_{-,r}^2$. (Before equation (16), Renaud also states that equation (16) 
follows since  $\phi_{+,r}=0$ even  in the presence of gravitational interactions. However,
to justify the use of the on-shell 
property it is in any case necessary to appeal to the perturbative argument he
gives below equation (17), according to which at lowest order the gravitational interactions 
are neglected. The relevance of the fact that 
 $\phi_{+,r}=0$ {\it even in the presence of gravitational interactions} is therefore not clear to me.)}

The action for the matter-gravity system on the background 
$g_0$ takes the form
\beq
S =S^\phi + S^g\,,
\eeq
where $S^g$ is the Einstein-Hilbert action for the metric $g=g_0+h$. 
The strongest 
gravitational interaction between the in and out modes of $\phi$ is produced by the 
$f\phi_r^2$ term in \eqref{Sphi},
\beq\label{Sint}
S_{\rm int} = G\int dv\, dr\, (\m/r)\,  \phi_r^2\,.
\eeq
The gravitational constraint equations fix
$\m$ and $\psi$  in terms of $\phi$ (and the background metric).
To simplify the computation Renaud
includes here only
the lowest order contribution in powers of the gravitational constant $G$.
To solve for $\mu$ and $\psi$ at this order one could expand 
$S^h$ to quadratic order in $h$ and integrate out $h$ from the resulting action,
which amounts to solving the linearized Einstein equation for  
$\mu$ and $\psi$. Since $\psi$ does not appear in the scalar field action,
it is determined in terms of $\mu$ and the background, and is in any case not
relevant for the gravitational self-interaction of $\phi$, 
so we may restrict attention to 
solving for $\m$.

Instead of working with a symmetry-reduced Einstein-Hilbert action, 
we can just use the standard, tensorial Einstein equation. 
The linearized Einstein
tensor component $G^{(1)}_{vv}$ with the metric \eqref{ds2} is 
\beq
G^{(1)}_{vv} = \frac{2}{r^2}\,G\m_v + \frac{f_0}{r^2}\,\Bigl[2G\m_r + \psi + r\psi_v + f_0r\psi_r\Bigr]\,.
\eeq
The  term in square brackets does not appear in Renaud's equation (17). 
Perhaps he neglected it 
because it is proportional to $f_0$, which is very small near the horizon 
where the interactions of interest are localized. 
However, in view of the $r$ derivative in $\m_r$, it is not clear to me that
it is subdominant. In particular, using the same partial derivative identity
as in \eqref{phirv}, we have $f_0\m_r = -2\mu_u|_v$, which is unsupressed
by $f_0$. On the other hand, this would receive contributions only from 
$\phi_-$, so would not contribute to the in-out mode interaction. In any case,
I will neglect it.

The linearized Einstein equation sets $G^{(1)}_{vv}$ equal to the metric-linearized 
part of $8\pi G T_{vv}$,
where the component $T_{vv}$ of the scalar field energy-momentum tensor is
\beq
T_{vv} = \chi_v^2+  2\pi r^2 \,f\,(\chi_v\chi_r + f\,\chi_r^2) \,.
\eeq
The second term on the right hand side also does not appear in Renaud's equation (17).
Again, perhaps he neglects it because it is proportional to $f$ \eqref{f},
but $f_0\chi_r = -2\chi_u|_v$, so it is not actually suppressed. However,
as with the above contribution to the Einstein tensor, it would involve only
the outgoing field, so would not contribute to the in-out interaction. 
Neglecting  it,
this component of 
the linearized
Einstein equation becomes simply
\beq\label{muv}
\m_v =  \phi_v^2
\eeq
after taking the definition \eqref{phi} into account.

\subsection{Gravitational interaction between in and out modes}
\label{sec:gravint}

 In general $\m$ depends
on both $v$ and $r$, so \eqref{muv} does not fully determine $\m$.
However, as explained above, in the near-horizon region the 
dominant contribution to the interaction comes from a contribution
to $\mu$ that depends only on the ingoing mode, which has no
$r$-dependence at fixed $v$. Renaud
neglects the $\phi_-$ dependence of $\mu$,
so that the solution to \eqref{muv} 
becomes
\beq\label{mu}
\m_+(v) := -\int_v^\infty dv'\, \phi_{+,v'}^2\,m
\eeq
where the subscript $+$ is Renaud's notation to indicate that this is 
the metric perturbation driven by $\phi_+$ flucutations. 
Since it depends only on $v$, $\m_+$ is nothing but a $v$-dependent
fluctuation in the mass of the line element \eqref{ds2}.
I have chosen the integration constant (operator) so that 
$\m_+(\infty)=0$, that is, the metric perturbation vanishes at future
null infinity.\footnote{Renaud chose instead in his equation (17) to
integrate from $0$ to $v$, in effect setting the perturbation to 
zero at the time of the shell collapse that forms the black hole. 
This does not seem to me well justified. Moreover, it leads to 
a divergent contribution to $\la\m_+(v)\m_+(v')\ra$ that he neglects in
his equation (38).}

\begin{figure}
     \centering 
     \includegraphics[width=0.4\linewidth]{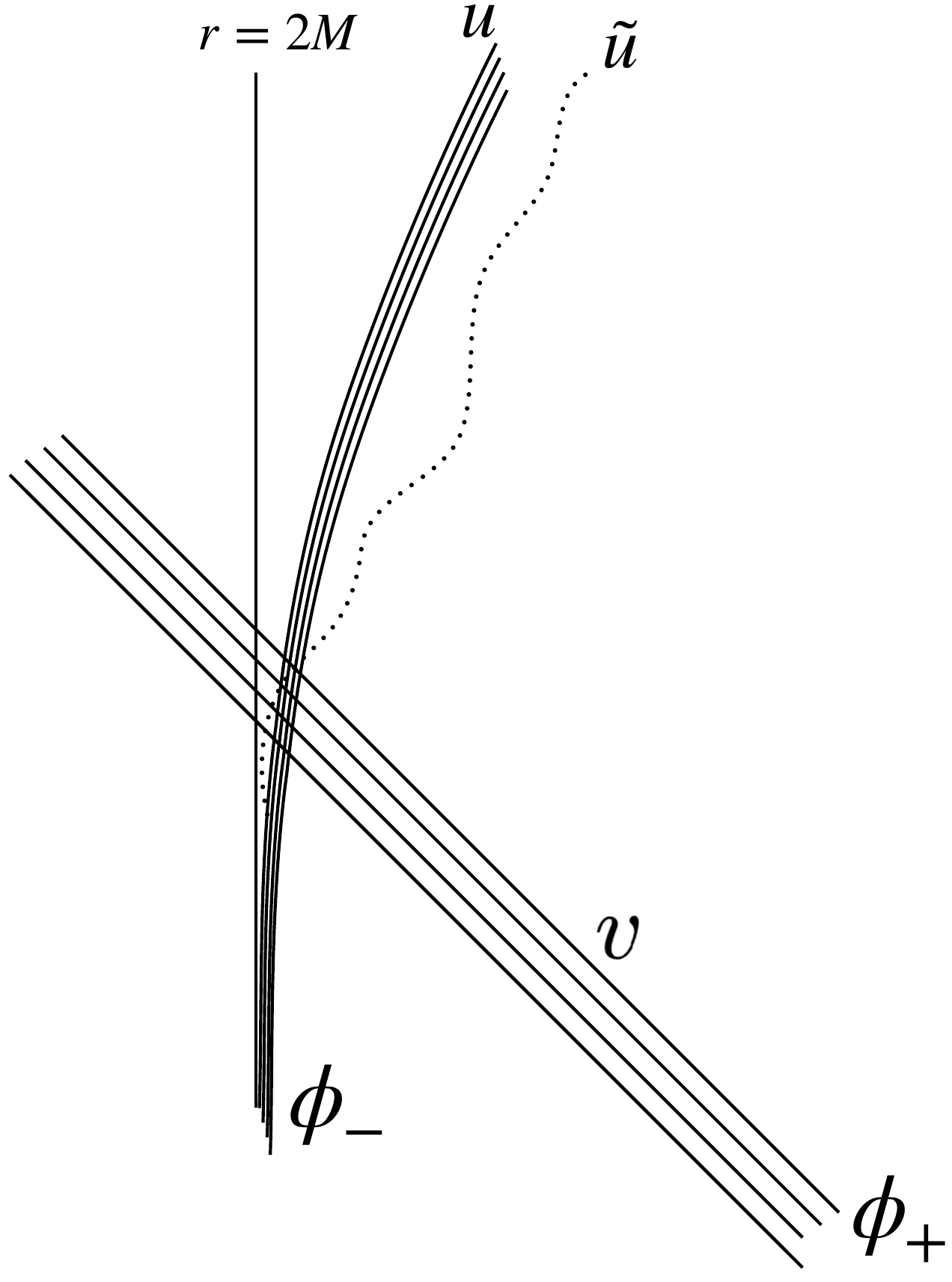}
     \caption{Spacetime sketch of the near-horizon region of a black hole,
     with some constant phase lines for a pair of ingoing 
     and outgoing modes, $\phi_+(v)$ and $\phi_-(u)$.
      The curve labeled $\tu$ is a sample null curve of the 
      metric with dispersion driven by the $\phi_+$ vacuum fluctuations.}
     \label{fig:figure}
 \end{figure}

Since $\mu_+$ is independent of $\phi_-$, 
equation.\ \eqref{nonlin}
becomes a linear equation for $\phi_-$.
Since $2\partial_v + f\partial_r$ is an outgoing null vector of the line element \eqref{ds2}, 
the solutions are constant on the modified outgoing 
null curves, $\phi_-(v,r) = \phi_-(\tu)$, 
where $\tu(v,r)$ is the retarded time in the presence of the 
gravitational fluctuations caused by $\phi_+$. 
This fluctuating retarded time 
satisfies the same equation \eqref{nonlin} as does the outgoing mode,
\beq\label{tu}
2\tu_v + \left(f_0(r)- \frac{2G\m(v,u)}{r}\right)\tu_r = 0\,.
\eeq
Renaud imposes the teleological boundary condition that in the asymptotic future 
$\tu$ becomes the retarded time coordinate in flat spacetime,
i.e., 
$\tu(v,r)\rightarrow v - 2r$ as $v,r\rightarrow\infty$
\eqref{u}. The reason for this choice is presumably that the 
quantum state far from the horizon is well described in the semiclassical 
approximation by perturbative excitation of the flat space vacuum.
(A spacetime sketch of the curves at play is shown in Fig.~\ref{fig:figure}.)
The solution to $\eqref{tu}$ is characterized, to first order in $G$, by
the variation 
\beq
\d u :=\tu - u\,,
\eeq
which satisfies
\beq
(\d u)_v\big|_{u} = G\,\frac{\mu(v,u)}{r}u_{r}|_v = -G\,\frac{2\mu(v,u)}{r(v,u) - 2M}
\eeq
(where \eqref{u} and \eqref{v|u} have been used). The $u$ fluctuation
is thus given to first order in the gravitational interaction by
\beq\label{du}
\d u(v,u) = G\int_v^\infty dv'\,\frac{2\mu(v',u)}{r(v',u) - 2M}\, .
\eeq
Note that the effect produced by the fluctuations of $\m$
is amplified near the horizon, due to the small denominator in \eqref{du}.
When including only the $\phi_+$ dependence of $\mu$, $\mu(v',u)$ in
the integrand of \eqref{du} is replaced by $\mu_+(v')$ \eqref{mu}.

In the quantum theory we promote \eqref{phi-} to the Heisenberg equation for the field 
operator (modulo operator ordering that is important in the exact treatment but will be
irrelevant in the perturbative treatment we are pursuing). 
When retaining only the $\phi_+$ dependence of $\mu$, 
the outgoing field operator $\phi_-(\tu)=
\phi_-(u + \d u)$ becomes an operator-valued
function of an operator-valued argument, with $\d u$  given by 
\eqref{du}, which depends on the field operator
$\phi$ via  \eqref{mu}.\footnote{This function of an operator 
can be made more concrete by expressing it in terms of the Fourier 
transform $\widetilde{\phi}(k)$,
as $\phi_-(u + \d u) = \int \frac{dk}{2\pi} \widetilde{\phi}(k)e^{ik(u + \d u)}$.}
In the approximations and simplifications adopted by Renaud, 
the fluctuation  $\d u$ is entirely induced by the fluctuations of 
the ingoing mode $\phi_+(v)$. 
This generates entanglement between the in and out modes which,
as we'll see, is responsible for decohering the $\phi_-\phi_-$ correlator.

The quantum fluctuations of $\d u$ are characterized by the 
variance, $\la\d u(v,u) \,\d u(v,u)\ra_{\rm C}$. 
To compute  this using \eqref{du}
we first need to compute the connected part of the correlator $\la \m_+(v)\m_+(v')\ra$
of the metric fluctuations,   
\beq
\la \m_+(v)\m_+(v')\ra_{\rm C}:=\la \m_+(v)\m_+(v')\ra-\la \m_+(v)\ra\la\m_+(v')\ra\,.
\eeq
The metric fluctuation $\m_+(v)$ is determined by an integral over
$\phi_{+,v}^2$ \eqref{mu}. Since $\d u$ already has a factor of $G$,
we only require the free field correlator of $\phi_{+,v}^2$, which is
determined by the two-point function 
\beq
\la \phi_+(v_1)\phi_+(v_2)\ra = -\frac{1}{4\pi}\ln(v_1 - v_2 - i\e) + \mbox{constant}\,. 
\eeq
It follows that\footnote{Renaud has the coefficient $1/16\pi^2$ in his equation (37),
but I think the two equivalent Wick contractions result in a factor of 2. In any case, 
the precise numerical factor is not relevant for his analysis.}
\beq
\la (\phi_{+,{v_1}})^2 (\phi_{+,{v_2}})^2\ra_{\rm C} = \frac{1}{8\pi^2}\frac{1}{(v_1-v_2-i\e)^4}\,,
\eeq
so we have
\begin{align}
\la \m_+(v)\m_+(v')\ra_{\rm C}
&=\frac{1}{8\pi^2}\int_v^\infty dv_1\int_{v'}^\infty dv_2\,\frac{1}{(v_1-v_2-i\e)^4}\nn\\
&= -\frac{1}{48\pi^2} \frac{1}{(v-v'-i\e)^2}\nn\\
&= \frac{1}{48\pi^2}\int_0^\infty d\o \, \o \, e^{-i\o(v - v'-i\e)} \,.\label{mpmp}
\end{align}
(The second equality is obtained 
by writing $(v-v'-i\e)^{-4} = -\frac16\partial_v\partial_{v'}(v-v'-i\e)^{-2}$.)
With this correlator in hand, we are now in a position to compute the 
variance of $\d u$ using \eqref{du}, which we postpone to the next section. 

To conclude this subsection, it is worth noting that, while I discussed the 
interaction using the Heisenberg picture, Renaud phrased the physics largely
in terms of the action. In particular, at first order in $G$, he argues that
the interaction action \eqref{Sint} is 
\beq\label{Sint2}
S_{\rm int} = G\int_0^\infty dr\int_0^\infty dv \int_0^v dv'\, r^{-1}\, \phi_{+,v'}^2\phi_{-,r}^2\,.
\eeq
Having ``integrated out'' the linearized
metric perturbation, the induced gravitational
interaction between ingoing and outgoing modes 
has the form $T_+\cdot{\cal D}^{-1}\cdot T_-$,
where $T_\pm$ is the energy-momentum tensor of the ingoing/outgoing modes,
and ${\cal D}$ is the ``kinetic'' operator for the metric perturbation.
In the present context, which is reduced to spherical symmetry,
and restricted to first order in $G$,
this yields \eqref{Sint2}.
It is of key importance that 
the interaction diverges at the horizon, in the following sense. 
If
$\phi_+\sim e^{i\o v}$ and $\phi_-\sim e^{i\l u}$, so that $\o$ and $\l$ are the asymptotic black hole rest frame frequencies of both modes, 
then $\phi_{+,v}\sim \o \phi_+$, and (using \eqref{r|v}) 
\beq\label{1/f}
\phi_{-,r}|_v= -2 f_0^{-1}\phi_{-,u}|_v\sim  f_0^{-1}\l \phi_{-}|_v\,.
\eeq
The interaction density
is thus proportional to $(\o\l/f_0)^2$, which diverges at the (unperturbed) horizon.

\section{Quenching of the 
two-point correlation function}

\begin{figure}
     \centering 
     \includegraphics[width=0.4\linewidth]{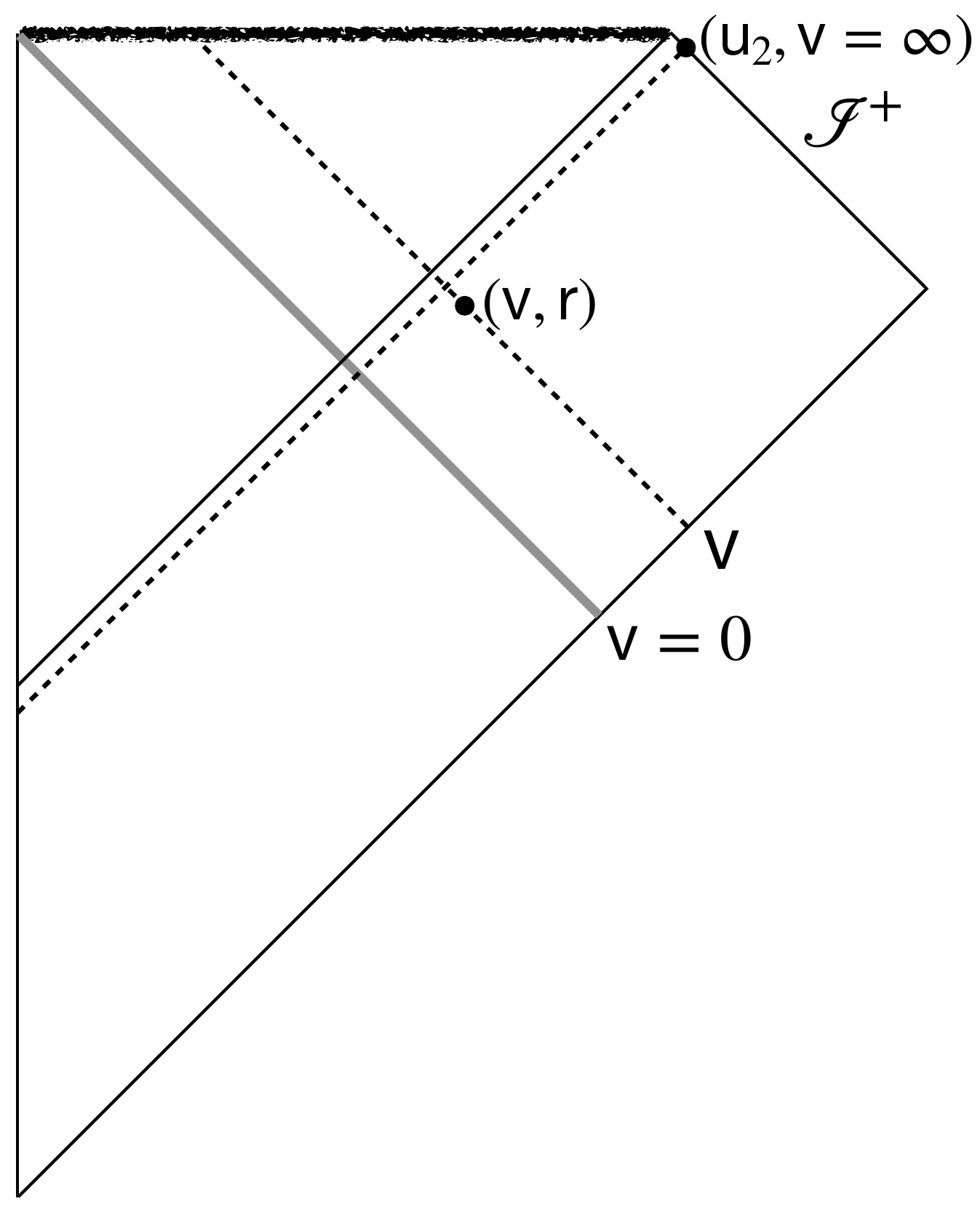}
     \caption{Penrose diagram of black hole formed from collapse of a spherical 
     null shell (thick grey line)
     at the ingoing null coordinate $v=0$. 
     The correlation function of interest is located at the points 
     labeled by coordinates $(v,r)$ and $(u_2, v = \infty)$, where $r$ is the
     spherical radius and $u$ is the outgoing null coordinate. 
     ${\cal I}^+$ is future null infinity,
    which coincides with $v=\infty$, and $u\to\infty$ at the future endpoint of ${\cal I}^+$.}
     \label{fig:BHdiagram}
 \end{figure}

The quantity to be studied for diagnosis of the 
quenching of trans-Planckian correlations 
is the two-point correlation function for the outgoing field,
\beq
\la\phi_-(x_2)\phi_-(x_1)\ra\,.
\eeq
The expectation value is to be taken in the in-vacuum state.
The points are specified 
as follows (see Fig.~\ref{fig:BHdiagram}):  $x_1$ is located by the spherical radius $r$ and the 
advanced time $v$ of a radial null geodesic at 
${\cal I}^-$, and the point $x_2$ sits on ${\cal I}^+$ at the retarded time $u_2$. 
Since the gauge diffeomorphisms do not act at ${\cal I}^\pm$, and
$r$ is intrinsically defined (in the spherically truncated context),
the correlation function is gauge-invariant. 
In the absence of the gravitational interaction, i.e., at zeroth order in $G$, it is proportional to $\ln\bigl(u_2-u(v,r)-i\e\bigr)$.
As discussed above, the effect of the gravitational interaction is to turn $u(v,r)$ into an operator $\tu(v,r)$ which acts on the Hilbert space of ingoing modes. 
The Fourier transform of the $u_2$ dependence 
of the two-point function is proportional to $e^{-i\l \tu(v,r)}/\l$, so 
the expectation value 
\beq\label{ft}
\la \widetilde{\phi}_-(v=\infty, \l)\phi_-(v,r)\ra \propto \la e^{-i\l \tu(v,r)}\ra
\eeq
characterizes the fate of 
the two-point function in the presence of the metric fluctuations driven by 
the quantum fluctuations of the ingoing modes. The challenge now
is to compute this quantity including nonperturbative effects in $G$.

As discussed in section \ref{role}, Renaud introduces at this point 
a large number $N$ of identical scalar fields.
An expansion in $1/N$ at fixed $GN$ then
allows him to exponentiate the effect of the gravitational
interaction without the need to solve the equations
of motion beyond linear order in $G$.
It also 
allows him to import from the 
Schwinger-Keldysh closed time path 
formalism \cite{Hu:2020luk} a key result:
when the environment of a quantum system is ``Gaussian'', i.e., 
governed by a quadratic action, the effect of the environment on the rest of the 
system is captured by the so-called noise and dissipation kernels, and 
the effect of
the noise kernel is identical to that produced by a Gaussian distribution of 
noise~\cite{Feynman:1963fq}.

The ingoing modes 
drive fluctuations of the metric variable $\m_+$, 
which produce an effective noisy environment 
for the outgoing modes. 
(Renaud focuses on the effect of the noise alone,
presumably since it is easily captured with the stochastic 
method, but a more complete treatment would also include
the dissipation.) 
To implement this scheme one needs to 
solve \eqref{tu} for the fluctuating retarded time 
at first order in $G$, and then compute the stochastic approximation to the
expectation value \eqref{ft}, which will involve all orders of $G$.

With the retarded time expanded around the background value as $\tu = u + \d  u$,
the expectation value \eqref{ft} takes the form 
\beq\label{fts0}
\la e^{-i\l \tu(v,r)}\ra =   e^{-i\l u(v,r)} \la e^{-i\l \d u(v,r)}\ra\,.
\eeq
In the large $N$ approximation, taking into account just the noise kernel, 
the quantum expectation value is replaced 
by the stochastic average with respect to a Gaussian probability distribution, 
\beq
P(\d u) = \frac{1}{\sqrt{2\pi \s^2}}e^{-\d u^2/2\s^2}
\eeq
where the variance $\s^2 = \la \d u\,\d u\ra_{\rm C}$ is the quantum expectation value,
and its position dependence is suppressed.
This yields the approximation for the Fourier transform of the two-point function
\beq\label{fts}
\la \widetilde{\phi}_-(v=\infty, \l)\phi_-(v,r)\ra \sim e^{-i\l u(v,r)} e^{-\l^2 \s^2/2}\,.
\eeq
It just remains to compute $\s^2$.

The variance $\s^2$ receives equal contributions from the $N$ scalars.
Using \eqref{du} for the $\d u$ integral and \eqref{mpmp} for the $\m_+$ correlator 
we find\footnote{Note that $\s^2$  
depends on only the symmetric, real,  part of 
$\la\mu_+(v_1)\mu_+(v_2)\ra_{\rm C}$.}  
\begin{align}
\s^2 &= \la\d u(v,u) \,\d u(v,u)\ra_{\rm C}\nn\\
&=
4G^2N\int_v^\infty dv_1\int_v^\infty dv_2 \,
\frac{\la\mu_+(v_1)\mu_+(v_2)\ra_{\rm C}}{[r(v_1,u) - 2M][r(v_2,u) - 2M]}\nn\\
&\approx\frac{4G^2Ne^2}{(4M^2)(48\pi^2)}\int_v^\infty dv_1\int_v^\infty dv_2 \,e^{\k(u-v_1)}e^{\k(u-v_2)}\int_0^\infty d\o\;\o \,e^{-i\o(v_1-v_2-i\e)}\nn\\
&\approx\frac{G^2N\k^2}{(r/2M-1)^2}\,\frac{1}{3\pi^2}\int_0^\infty d\o\,\frac{\o}{\o^2+\k^2}e^{-\o\e}\label{oint}\,.
\end{align}
Here $\k:=1/4M$ is the surface gravity of the background black hole.
The approximate equalities use \eqref{u} together with the near-horizon
approximation $e^{r/2M}\approx e^1$. 
The integrand of the $\o$ integral in \eqref{oint} peaks at $\o=\k$ for $\e=0$, but
in the limit $\e\rightarrow0$ the integral diverges as $\g-\ln(\k\e)$,
where $\g$ is Euler's constant.  If we introduce 
a UV frequency cutoff $\Lambda=1/\e\gg\k$, the integral becomes
approximately $\ln(\Lambda/\k)$.\footnote{Renaud did not 
include $\epsilon$ in the $\o$ integral, but rather just introduced 
$\Lambda$ as the upper limit of integration.}
As mentioned in section \eqref{role}, 
the UV divergence in this crude model
may result from the fact that the flat spacetime gravitational vacuum interactions
have not been suitably subtracted. Nevertheless, 
the logarithmic dependence on the cutoff is very weak, 
so perhaps a correct physical lesson can be drawn.

The standard deviation $\s$ in \eqref{oint} 
depends on $v$ and $u$ only via $r$, and can be written as
\beq\label{sigma}
\s = \frac{\bar \s}{r/2M-1},\qquad \bar\s := G\k \sqrt{\frac{N\ln(\Lambda/\k)}{3\pi^2}}\,.
\eeq
According to \eqref{fts}, the Fourier component of the two-point correlation function 
at Killing frequency $\l$ will decay to zero as a Gaussian when $\l\s\gg1$, i.e., when 
\beq\label{decay}
r/2M-1\ll \l \bar\s\,.
\eeq
The two-point correlation function itself is thus modified from 
$\ln\bigl(u_2-u(v,r)-i\e\bigr)$, which is singular at $u_2=u(v,r)$, to a function that is regular in the UV.
In the near horizon limit we have $r/2M-1\approx \k^2\ell^2$,
where $\ell$ is the radial proper distance to the horizon (i.e., to the bifurcation
surface in the maximal extension of the Schwarzschild metric) 
on a ``static slice'' (normal to the Killing vector). The decay condition
\eqref{decay}
may thus be expressed as  $\ell^2 \ll \l\bar\s/\k^2$.
Using \eqref{sigma}, with the square root taken as $O(\sqrt{N})$, 
the decay condition  
becomes 
\beq\label{ell}
\ell\ll \ell_c:=\sqrt[4]{N}\sqrt{\l/\k}\; \ell_{\rm P}\,,
\eeq
where $\ell_{\rm P}=\sqrt{\hbar G/c^3}$ is
the Planck length. At a frequency equal to that of the 
Hawking temperature ($T_{\rm H} = \k/2\pi$), the correlator thus dies off when the point $x_1$ 
is within $\sim \sqrt[4]{N}$ Planck lengths of the horizon, measured on a static slice.

\subsection{Validity regime questions}

The Gaussian $e^{-\l^2\s^2/2}$ that damps the Fourier transform of the two-point function
can be expanded as a power series in $\s^2$, which according to \eqref{oint} is proportional 
to $G^2N$. Since $GN$ is held fixed, this is a series in powers of $1/N$, which is thus formally
convergent, despite containing arbitrarily high orders in $G$. This series is what
Renaud pointed to as the first contribution beyond the semiclassical approximation, 
and it accounts for the title of his paper. 
One should ask, nevertheless, what is the relative contribution of the series in powers of $G^m N$, with $m>2$?
Term by term, these series will be smaller than the $G^2N$ series provided the dimensionless
quantity formed with $G$ is small compared to 1. If the quenching effect kicks in at a length scale
above the Planck scale, then presumably that dimensionless quantity is small compared to 1.
According to \eqref{ell}, that seems indeed to be the case, provided that the
divergent quantity $\ln(\Lambda/\k)$ that was treated as $O(1)$ in \eqref{ell} 
would in fact not appear in a correctly renormalized treatment.

It is encouraging, for maintaining validity of the approximations,
that the critical length $\ell_c$ \eqref{ell}
is parametrically larger than the Planck length, but 
we should enquire as to the meaning of that length, and as to whether the 
Planck length is the right length to compare to.
First, while
$\ell$  measures proper distance to the bifurcation surface 
on a static slice of the Schwarzschild spacetime,
there is 
no bifurcation surface in a spacetime with collapse to a black hole. Nevertheless, shortly after the collapse, $\ell$ is
also very close to the proper distance to the surface of the collapsing 
matter: equation \eqref{ratio2} shows that 
the ``missing length'' $\ell_1$ divided by the full length $\ell_2$ 
to the would-be bifurcation surface decreases exponentially as $e^{-\k\D v}$.
Thus $\ell$ is a meaningful length scale in the collapse spacetime. 

But even if $\ell_c > \ell_{\text{P}}$, 
would that be long enough for the large $N$ perturbation theory to be justified? 
Since we have assumed a large number of matter fields $N$, holding $GN$ fixed, 
perhaps the condition for perturbation theory not to fail is not $\ell_c>\ell_{\rm P}$
but rather $\ell_c>\sqrt{N}\ell_{\rm P}$, which is equivalent to 
$\l > \sqrt{N}\k$. This would still permit the conclusion 
that the singularity of the two-point function is quenched, 
but it may be quenched at a larger value, since the calculation would only 
imply that 
Fourier components with $\l > \sqrt{N}\k$ would be damped.

\section{Discussion}
The \tP problem has several facets: the origin of the outgoing black hole modes
and the Hawking radiation they carry, 
the divergent contribution of quantum field entanglement to black hole entropy, 
and the origin of degrees of freedom in an expanding universe. 
These are not problems at the level of effective field theory,
but they seem problematic when one digs for a deeper level of understanding,
in part because of our incomplete understanding of quantum gravity in general. 
A major motive for  attempting to tackle them is that they 
may provide guidance in the quest for understanding quantum gravity. 

This article focused on Renaud Parentani's analysis of 
a black hole in quantum gravity, which aimed to extract
consequences of the gravitational interaction of
ingoing and outgoing modes {\it in their ground states}.
The model is similar to earlier work \cite{tHooft:1984kcu,tHooft:1986vqu, tHooft:1994tah, tHooft:1996rdg, Kiem:1995iy}
which also studied
the gravitational interaction of these modes, but with
a key difference in aim. In those works, the focus was on
the black hole S-matrix, in particular, the sensitivity of that
S-matrix to the addition of an infalling particle. That
work was motivated by the black hole information paradox,
the question being whether these interactions 
could play a role in preservation of unitarity. This question 
has persisted to recent times, as seen for example 
in~\cite{Shenker:2013pqa, Polchinski:2015cea}.
In contrast, Renaud's study was focused on the vacuum, 
and designed to explore the consequences of quantum geometry 
fluctuations for the \tP problem. Nevertheless, the S-matrix 
(at least in standard interacting quantum field theory)
is closely related to vacuum correlation functions, so the two
aspects of near horizon physics may be similarly related.
Moreover, the discontinuous shift of retarded time
in response to the gravitational effect of 
an incoming null shock, 
studied in \cite{tHooft:1984kcu,Shenker:2013pqa},
is an abrupt limit of Renaud's
fluctuation $\d u$ \eqref{du} of retarded time 
in response to the quantum gravitational
effect of the incoming vacuum modes.

Renaud aimed to make some progress on this
very difficult problem by admitting drastic simplifications:
the spherical truncation, the neglect of subleading and other 
effects not central to the 
in-out-interaction question, and the use of a large $N$
approximation.
Within that framework, he uncovered what looks like it might 
point to a quantum gravity mechanism that  quenches
\tP near-horizon correlations. 
How valid it is remains to be
seen, but in my view this approach deserves to be developed
and further explored. 

It might be fruitful to study the 
problem in a two-dimensional theory like JT gravity where,
rather than relying on a severe truncation of a higher-dimensional 
theory, one could come closer to a complete analysis. 
On the other hand, 
the \tP problem might have a different status in  
two dimensions.
Perhaps a model similar to Renaud's could be implemented
in four-dimensional spacetime without the spherical truncation,
using the well-developed methods of the influence functional 
as applied to gravity~\cite{Hu:2020luk}. In fact, those methods have
been applied in flat spacetime~\cite{Hu:2004gf}, 
to fields outside black holes~\cite{Hu:2007tq},
and in de Sitter spacetime~\cite{Frob:2014cza}, but not to the sort of 
near-horizon correlator, including nonperturbative effects in $G$, that Renaud targets.
Furthermore, it would be important to determine whether higher derivative 
interactions in the effective theory might dominate
lower derivative ones, since each derivative for outgoing modes brings in
another factor of $1/f_0$ \eqref{phirv}.

Another possible avenue to explore would be the case of a black hole in 
asymptotically anti-de Sitter spacetime, where AdS/CFT duality
could be exploited. 
As a bulk point approaches the horizon in that setting, 
its boundary dual is a smearing of local operators over a region
that extends to infinite time. Its correlation with 
operators localized at its future light cone cut on the boundary
therefore decays exponentially on a thermal time scale $T^{-1}$,
or perhaps on the scrambling time scale $T^{-1}\log S$, 
(until the very late time $\sim S/T$ determined by the 
finiteness of the CFT entropy $S$ at the temperature $T$) \cite{Kabat:2014kfa}.
Might that thermal temporal decay in the boundary dual 
be related to the near-horizon decay
that Renaud found? Studies of the effect of shockwave perturbations
of the vacuum \cite{tHooft:1984kcu,Shenker:2013pqa} 
found scrambling of S-matrix elements or cross-horizon
correlations, which does seem ``morally'' related to Renaud's finding for
vacuum correlators. 
In the AdS/CFT context it is also worth mentioning the salient fact that
the finiteness of the black hole entropy, and thus a solution to the \tP problem, 
is at hand, since finiteness of the dual CFT entropy perfectly understood, even if
the detailed implications for near horizon physics remain currently out of reach.

What if refined calculations were to establish that indeed \tP near horizon correlations
are quenched by gravitational interaction governed by general relativity? 
What would be the broader implications for 
quantum gravity? It is already remarkable that the dynamics of general relativity
fits tongue in groove with the quantum field phenomenon of Hawking radiation, 
allowing for consistent black hole thermodynamics and the generalized second law.
This consistency was discovered initially at the semiclassical level 
on the gravity side. But such a deep unity must presumably extend to the level 
of the interacting quantum gravitational vacuum. 

If \tP physics were not masked 
by the gravitational quenching of correlations, then black hole horizons would 
have been windows into the \tP regime, which would likely have led to a breakdown
of the effective field theory description of gravity. 
Moreover, it would seemingly have led to an infinite black hole entropy,
which would not only disagree with the Bekenstein-Hawking entropy,
but would also invalidate the thermodynamic derivation of the Einstein
equation as a vacuum equation of state \cite{Jacobson:1995ab, Jacobson:2012yt}. 
The masking is thus
to be expected, and Renaud's calculation suggests that a mechanism
can already be discerned at the level of perturbative quantum gravity, provided 
that some nonperturbative consequences can be captured by a suitable, justified 
resummation like that of the large $N$ approximation.\footnote{A locally Lorentz-invariant 
mechanism to cloak the UV entanglement by gravitational 
back-reaction of the entangled pairs was proposed qualitatively in \cite{Jacobson:2012yt}.
That seems to be quite different from Renaud's mechanism, 
because it does not rely on the spacetime background being a black hole,
and is not (at least in any obvious way) 
related to interactions between ingoing and outgoing modes.}  
In fact, Renaud proposed in \cite{parentani2007beyond} that, 
in the setting with the nonspherical degrees of freedom included,
the length scale $\ell_c$ at which the near-horizon correlations are
quenched should also be the length scale at which the entanglement 
accounting for black hole entropy is cut off.  If this were the 
case, the scaling of the entropy with the number of 
fields $N$ might be canceled by the $N$ dependence of $\ell_c$.\footnote{Note that
the $N^{1/4}$ dependence of $\ell$ in the spherical reduction
result \eqref{ell} would {\it not} cancel, since the entanglement
entropy would scale as $N/\ell^{2}\propto N^{1/2}$ (in four spacetime dimensions).}

Where would this leave us regarding the other facets of
the \tP problem? Would it reveal how the outgoing modes are 
generated near a black hole horizon? And would it address 
the question of how new cis-Planckian modes emerge 
in an expanding universe? 
Of course I don't pretend to know the answers,
but allow me to speculate. 

It seems plausible that the answer to the first
of these questions is yes. In the presence of the strong gravitational
interactions, ingoing and outgoing modes are thoroughly mixed, and 
it is not implausible that the presence of a black hole horizon could
engender a net ``spectral flow'' of outgoing modes emerging from the near
horizon region. In fact, these words are quite reminiscent of 
the idea that a chiral diffeomorphism anomaly underlies the phenomenon
of Hawking radiation~\cite{Robinson:2005pd,Iso:2006wa,Sou:2025ozb}.
But I have not yet managed to 
fully understand the line of thought in those cited papers, and in particular
the precise role that ``mode creation'' versus ``mode excitation'' plays. 

As to the second question, it seems that gravitational interaction in
the fixed Hilbert space of an effective theory, no matter how strong, 
is powerless to create new degrees of freedom. But even that is not
to say that gravity does not have a role to play in that mysterious
process. It may just be that nonperturbative quantum gravity 
in an expanding universe does not admit a fixed Hilbert space
description. There are certainly reasons to suspect that is the case.
In a recent talk~\cite{pirsa_PIRSA:25060008}
I  sketched a way that such ``vacuogenesis'' might work, 
involving the ``wave function of the universe''.

%\par
%\section*{Data Availability}
%This work does not consider any new data.
\section*{Acknowledgments}
I am deeply grateful to Renaud Parentani for many years of 
discussion related to black hole physics and quantum gravity. I 
thank Daniel Kabat, Pranav Pulakkat, Albert Roura, Sergey Sibiryakov, and Yunfei Wang
for useful discussions and suggestions related to the topic of this paper and its presentation.
This research was supported in part by 
NSF grant PHY2309634.

\appendix
 
\section{Schwarzschild, Eddington-Finkelstein and double-null coordinates}
\label{A}
The angular part of $ds^2$ is $r^2 d\O^2$ in both coordinate systems, so I omit it.
For the background black hole metric we have
\beq
ds^2 = -f_0 \,dt^2  + f_0^{-1}\,dr^2 = -f_0\, dv\, du = -f_0 \,dv^2 + 2\, dv\, dr
\eeq
where $f_0=1-2M/r$, and 
\beq 
v = t + r_*,\quad u = t-r_*\,, \quad dr/dr_* = f_0\,, \quad r_* = r + 2M\ln(r/2M-1)\,,
\eeq
so in particular
\beq\label{u}
u = v -2r_*\,,\qquad r_* = (v-u)/2\,, \qquad r/2M-1= e^{-r/2M}e^{(v-u)/4M}\,.
\eeq

The relations between partial derivatives with respect to the 
$(v,u)$ coordinates and the $(v,r)$ coordinates are
\beq\label{v|u}
\partial_v|_{u} = \partial_v|_{r} + \half f_0\, \partial_r|_{v}\,, \qquad \partial_u|_{v} = -\half f_0\partial_r|_{v}\,,
\eeq
and 
\beq\label{r|v}
\partial_v|_{r} = \partial_v|_{u} +  \partial_u|_{v}\,,\qquad \partial_r|_{v} = -2f_0^{-1}\partial_u|_{v}\,.
\eeq

The $(v,r)$ coordinates of two points on 
a constant Schwarzschild time slice ($t = \mbox{const}$)
satisfy $v_1-r_{*1} = v_2 - r_{*2}$, 
which implies that 
\beq\label{ratio}
\frac{r_1 - 2M}{r_2-2M} = e^{(-\D v+\D r)/2M}\,,
\eeq
with $\D v = v_2-v_1$ and $\D r = r_2-r_1$.
If both points are near the horizon we have $\D r\ll 2M$, and 
$r-2M\approx \k^2\ell^2$, where $\k = 1/4M$ is the surface gravity and
$\ell$ is the proper distance to the horizon on a constant $t$ slice.
In the near-horizon case 
\eqref{ratio} thus implies that
\beq\label{ratio2}
\ell_1/\ell_2\approx e^{-\k \D v}\,.
\eeq
\bibliographystyle{JHEPmod}
\bibliography{beyond}
 
\end{document}